\documentstyle[osa,twocolumn,prl,aps,graphicx]{revtex}
\begin{document}
\newcommand{\beq}{\begin{equation}} 
\newcommand{\eeq}{\end{equation}}
\newcommand{\beqa}{\begin{eqnarray}}
\newcommand{\eeqa}{\end{eqnarray}}
\newcommand{\beqano}{\begin{eqnarray*}}
\newcommand{\eeqano}{\end{eqnarray*}}

\draft
\preprint{\footnotesize Revised \today}
\twocolumn[\hsize\textwidth\columnwidth\hsize\csname@twocolumnfalse%
\endcsname
\title{
Stripe Formation within SO(5) Theory}
\author{M. Veillette,$^1$ Ya.B. Bazaliy,$^2$ A.J Berlinsky,$^3$ and C. Kallin$^3$}
\address{$^1$Department of Physics, University of California, Santa Barbara, CA, 93106}
\address{$^2$Department of Physics, Stanford University, Stanford, CA, 94305}
\address{$^3$Department of Physics and Astronomy, McMaster University, Hamilton,
Ontario, Canada, L8S 4M1}
\date{\today}
\maketitle
\begin{abstract}
We study the formation of stripe order within the SO(5) theory of high-$T_{c}$ 
superconductivity.  Spin and charge modulations arise as a result 
of the competition between a local tendency to phase separate and the 
long-range Coulomb interaction. This frustrated phase separation leads
to hole-rich and hole-poor regions which are respectively superconducting 
and antiferromagnetic. A rich variety of microstructures ranging from droplet
and striped to inverted-droplet phases are stabilized, depending on the 
charge carrier concentration. We show that the SO(5) energy 
functional favors non-topological stripes.
\end{abstract}
]
One of the most striking features of the cuprates is the proximity
between antiferromagnetic (AF) and superconducting (SC) phases as
a function of doping. Recently it has been proposed that these two
phases are unified by an approximate SO(5) symmetry \cite{zhang1}. A
number of experimental consequences of this theory have been
worked out \cite{arovas,demler,burgess,bruus,goldbart}. Although 
SO(5) appears to be a natural framework for understanding the
cuprates, no experiment has unequivocally tested the fundamental
validity of the theory.  One of its most direct predictions is the
existence of a first-order transition from the AF to SC state as
the chemical potential $\mu$ is increased beyond a critical value.
However, this prediction is complicated by the fact that the doping
$x$ (not $\mu$) is the experimentally tunable parameter.
Experimentally, it is found that in the vicinity of the AF/SC
transition, the cuprates show an increased sensitivity to
disorder and inhomogeneity. In this Letter, we study this region of
the phase diagram in the presence of the long-range Coulomb
interaction within the SO(5) formalism and show how spatially
inhomogeneous states can emerge.

In the $T-\mu$ phase diagram of SO(5) theory, there is a
first-order line separating the AF and SC phases, across which the
charge carrier density $x$ jumps discontinuously. In the $T-x$ phase
diagram, this translates into a two-phase region where AF and SC
phases coexist.  Phase separation into hole-rich and hole-poor
regions was also noticed in studies of the $t-J$ model
\cite{emery1,dagotto}. However, as Emery and Kivelson \cite{emery}
argued rather successfully, the long-range Coulomb interaction between
charge carriers prevents macroscopic phase separation. The competition
between the local tendency toward phase separation and the long-range
Coulomb interaction leads to modulated domain structures at mesoscopic
scales \cite{low,stojkovic,castro,comment}.  In the SO(5) theory, the
hole-rich and hole-poor regions are respectively identified as having
a superconducting and antiferromagnetic character. The spin and charge
modulations of the system are interpreted as textures of the SO(5)
superspin as it rotates in SO(5) space.

There is considerable evidence for modulated microstructure in the
oxides.  Domain formation has been reported in La$_{2-x}$Sr$_x$CuO$_4$
(LSCO) in muon spin resonance \cite{borsa}, NMR and neutron
diffraction experiments \cite{suzuki}. Neutron scattering measurements
in La$_{1.6-x}$Nd$_{0.4}$Sr$_{x}$CuO$_4$ (LNSCO) provide direct
evidence for stripe ordering in which the phase of the AF order shifts
by $\pi$ across a domain wall \cite{tranquada}. Furthermore,
recent inelastic neutron scattering measurements in underdoped
YBa$_2$Cu$_3$O$_{7-x}$ (YBCO) \cite{dai} and ARPES measurements in
underdoped Bi$_2$Sr$_2$CaCu$_2$O$_{8-x}$ (BSCCO) \cite{shen} are not
inconsistent with a striped phase interpretation.

In the mean field approximation of the SO(5) theory one minimizes the
classical energy 
\begin{equation}\label{so5energy} 
H_1(n_a,p_a) =
\frac{1}{4} \sum_{ab} \frac{L^2_{ab}}{\chi_{ab}} + g(n_1^2 +n_5^2)
+\frac{\rho_s}{2} (\nabla \vec n)^2 + E_c 
\end{equation} with $L_{ab}=
n_a p_b - n_b p_a$ and constraints $n^2_a = 1$, $n_a p_a =
0$. $L_{ab}$ and $\vec{n}$ refer respectively to the SO(5) generators
of rotation and the 5-component superspin
$\vec{n}=(Re(\Delta),N_x,N_y,N_z,Im(\Delta))$ \cite{zhang1}. The last
term $E_c$ is the Coulomb energy. Since, in SO(5) theory, the hole
density is given by $L_{15}$, the charge density is
$\rho(r) = L_{15}(r) - e x$, 
where $ex$ is the charge of the neutralizing counterion charges which
are assumed to be static and homogeneously distributed with a
density~$x$. We conjecture then that, if SO(5) theory is still valid in
the presence of the long-range Coulomb interaction,/cite{coul} 
the Coulomb energy in
the mean field approximation will be given by
\begin{equation} E_c = \frac{1}{2} \int
\! \!  \int \rho(r) V_C(r-r') \rho(r')\; dr \;dr'\ . 
\end{equation}

It is important to emphasize that for homogeneous phases one recovers
the basic SO(5) model because the charge density vanishes exactly
making the Coulomb interaction irrelevant. Hence the influence of the
Coulomb term arises solely in the phase separation regime.  Second,
the assumption of immobile static counterions is known to fail for
La$_2$CuO$_{4+\delta}$ \cite{dabrowsky}. In this case, the oxygen ions
are mobile enough as to screen the charge inhomogeneities which leads
to macroscopic phase separation of superconducting and
antiferromagnetic domains. This last fact provides strong evidence for
the correctness of the SO(5) picture.

It can be proved that if $H_1$ is symmetric with respect to rotations within AF
and SC subspaces ($\chi_{15}=\chi_s,
\chi_{23}\!=\!\chi_{34}\!=\!\chi_{24}\!=\chi_a,
\chi_{12}\!=\!\chi_{13}\!=\!\chi_{14}\!=\!\chi_{23}\!=\!\chi_{24}\!=\chi_{\pi}$),
the minimal configuration has a form $\vec n =
(n_1,n_2,0,0,0)=(\cos\theta,\sin\theta,0,0,0)$, $\vec p =
(0,0,0,0,p_5)$ and the constraints are automatically satisfied if we
use the variables $(\theta,p_5)$.

With the addition of the long-range Coulomb interaction to the
Hamiltonian, the behavior of the system is no longer tractable
analytically and one must resort to numerical analysis. We
note that a classical spin Hamiltonian
\begin{equation} 
H_2 = J \sum_{\langle
  i,j \rangle} \vec{S}_i \cdot \vec{S}_j -2 K \sum_i (S_i^z)^2 
\label{xxz}
\end{equation} 
can be transformed into the local part of $H_1$ by a Haldane map (see
\cite{auerbach1} for details). Here $J>0$ and $K>0$ are known
functions of $\chi_{ab},g,\rho_s$.  
To match the whole expression (\ref{so5energy})
we add a term
\begin{equation}
V= \frac{1}{2} \sum_{i,j} \frac{(S_i^z -x)(S_j^z-x)}{\epsilon \left| \vec{r}_i - \vec{r}_j \right|}
\end{equation}
where $\epsilon$ is the dielectric constant of the material.

As emphasized earlier, since experiments are performed at constant carrier
concentration, the Hamiltonian $E=H_2 +V$ is subject to the doping
constraint $\langle S^z \rangle=x$. Numerically it is easier to study
$E$ than $H_1$ because the hole density is given explicitly by
$S^z$ rather than implicitly by $L_{15}$.  The properties of $E$ are
studied using Monte Carlo simulations.  In order to find the lowest
energy state of the system, we perform simulated annealing from
high-temperature. We assume an $N\times N$ 2-dimensional lattice where
$N$ can be up to 40 unit cells.

\begin{figure}[tbp]
\begin{center}
\label{phasedia}
\resizebox{7.0cm}{4.7cm}{\includegraphics{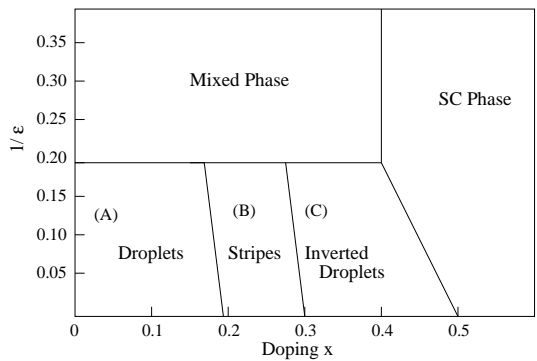}}
\vskip 0.1in
\caption{Phase diagram as a function of strength of the Coulomb 
interaction $1/\epsilon$ and the hole density $x$ for $K=0.4$ 
and $J=1$ which gives $x_c=0.5.$} 
\end{center}
\end{figure}

\begin{figure}[htbp]
\begin{center}
\begin{tabular}{c}
\resizebox{6.0cm}{4.0cm}{\rotatebox{0}{
         \includegraphics{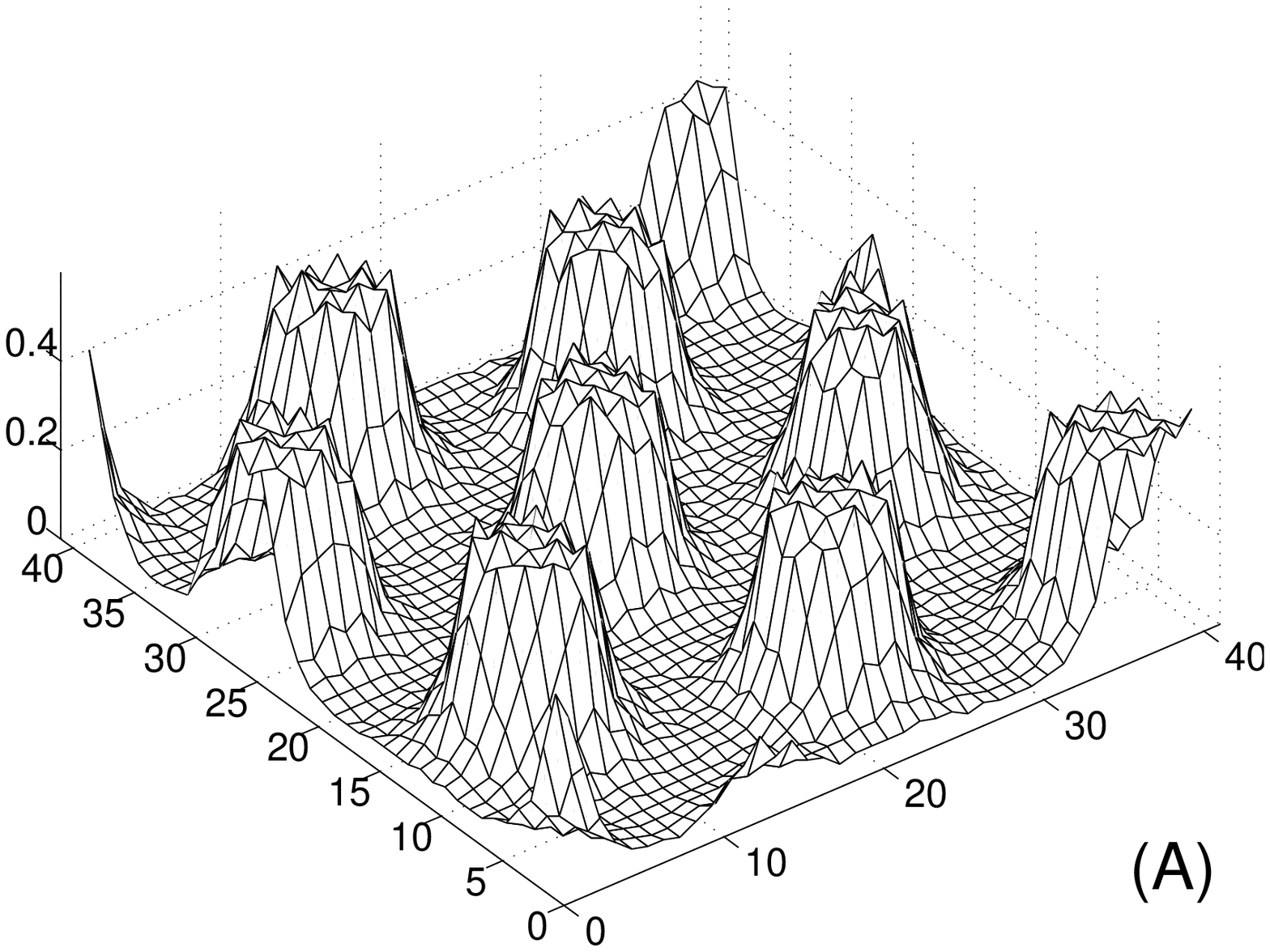}}} \\
\resizebox{6.0cm}{4.0cm}{\rotatebox{0}{
	\includegraphics{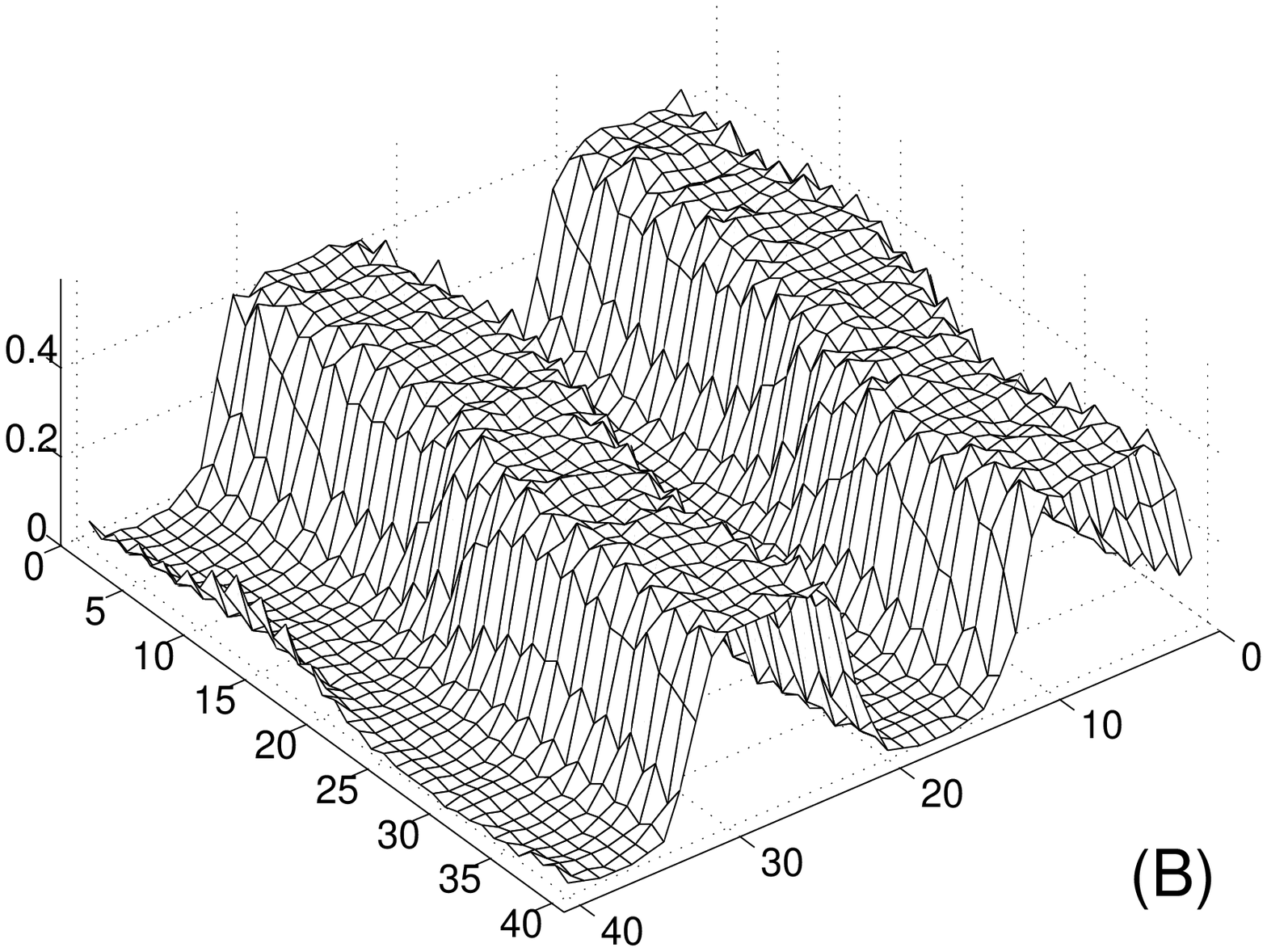}}} \\
 \resizebox{6.0cm}{4.0cm}{\rotatebox{0}{
	\includegraphics{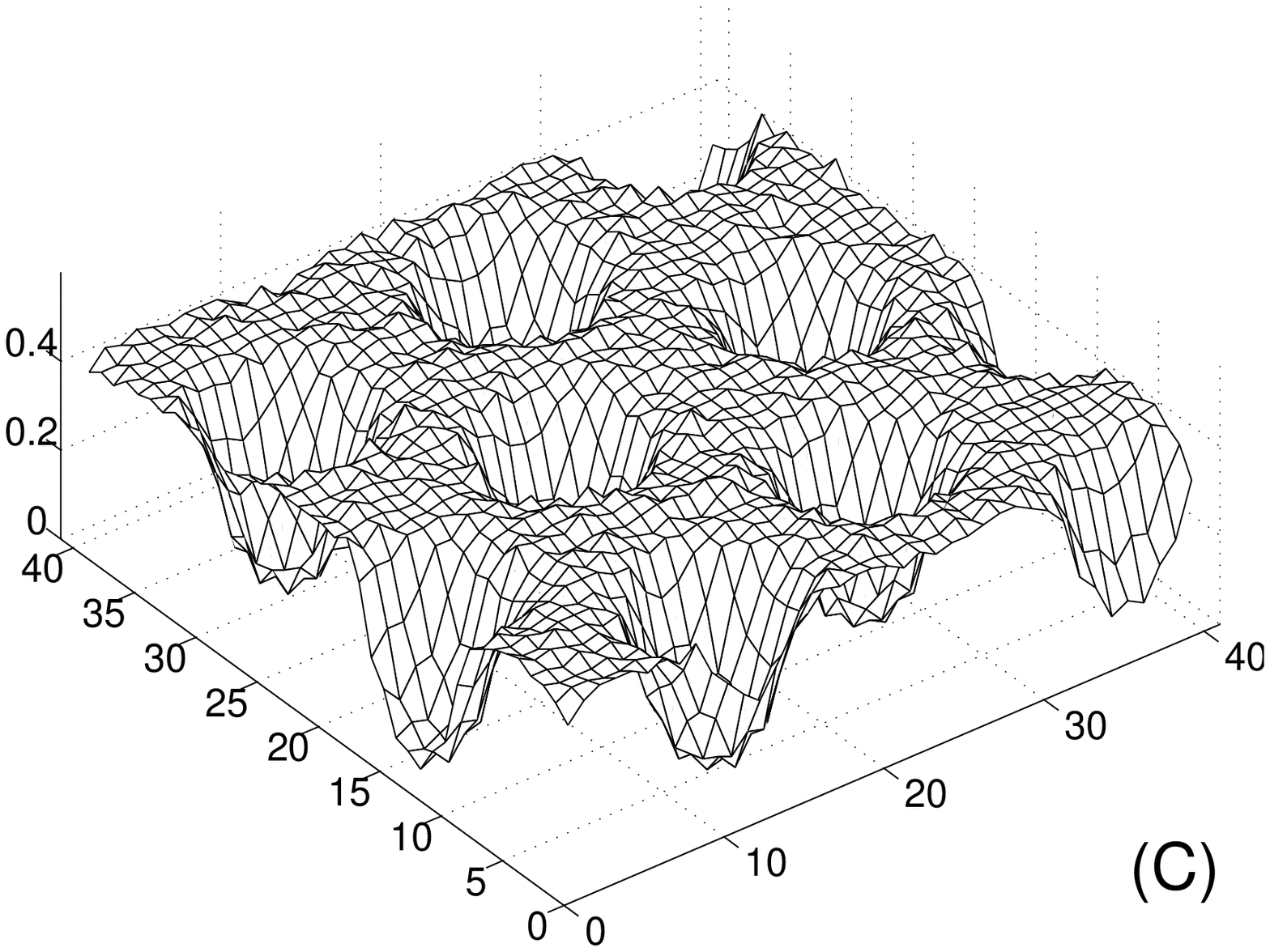}}} 
\end{tabular}
\end{center}
\caption{ Geometric phases as a function of doping: Hole density profile for the (A) Droplet phase ($x=0.12$), (B) Striped phase ($x=0.24$) and (C) Inverted-Droplet phase $x=0.35$. The parameters chosen are $J=1, K=0.4$ and $\epsilon=6$.}
\label{figure1}
\end{figure}

In the absence of the long-range Coulomb interaction, one can easily show
that the system phase separates for densities $x$ less than
$x_c=K/(2J-K)$.  The addition of the Coulomb term leads
to a rich variety of modulated structures, which are shown on the
phase diagram of Fig.\ref{phasedia}. For large dielectric constant,
three phases are found to be stabilized: a droplet phase made of SC
droplets embedded in an AF background, a striped phase of alternating
SC and AF stripes and an inverted-droplet phase where the droplets are
antiferromagnetic.  Our numerical solutions
(Fig.\ref{figure1}) show that the superspin stays in the AF or SC directions
inside the domains and changes only in the thin domain walls. The
structure represents a collection of solitons rather than a small
modulation of the direction of $\vec n$. The superconducting density
switches between $0$ and $x_c$ in AF and SC domains which means that
the superconducting area fraction $A_{sc}/A = x/x_c$ leading to a
linear relation between the superfluid density and the doping as seen
in some experiments\cite{bernhard}.  A simple physical argument for
the pattern shape can be given in terms of interface
energy\cite{seul}. As long as one of the phases (AF or SC) is in the
minority, the energy of the AF/SC interface predominates over the
Coulomb energy and circular domains are preferred as they minimize the
length of this interface.  However, the situation is reversed for $x
\approx  x_c/2$ where the repulsive interaction leads to
dipole formation which favors elongated domains such as stripes.

The striped phase is reminiscent of the domain structure observed in
LNSCO, though the rows of charge are superconducting in our model.
It is interesting to consider the superspin texture in the striped
phases, namely, the relative phase shift of every other stripe.
Numerically, it is found that the lowest energy states do not show any
winding of the superspin in space. Hence, the phase of the AF order
parameter does not shift by $\pi$ on crossing a SC stripe.  The same
results were obtained for a simulation on a spin ladder where all
configurations are necessarily one-dimensional.  

This absence of topological phase shift is in striking contrast to
experimental data \cite{tranquada}; however, it can be proved
analytically for the minimal periodic 1-D configuration of
(\ref{so5energy})  using the theorem of Pryadko et al. \cite{pryadko}. \\
{\em Theorem:} For a functional
\begin{eqnarray}
\nonumber
{\cal E} &=& \int (\frac{dv}{dr})^2 + E_{loc}(v^2(r),r) \; dr \\
&+& \int \!\!\!\int \rho(v^2(r),r) 
V(r-r^\prime) \rho(v^2(r^\prime),r^\prime) \; dr \; dr^\prime
\end{eqnarray}
of the function of one argument $v(r)$ the minimal configuration
under a constraint
$\int \rho(v^2(r),r) \; dr = 0$,
does not cross zero.

Applying it to (\ref{so5energy}) we note that as a function of
$\theta$ both $\rho$ and $H_1$ can be expanded in even powers around
the points $\theta = 0,\pi$ and thus $\theta_{min}(r)$ can not cross
these levels. 
In addition, let us perform a variable change
$(\theta,p_5)\to(\theta, q = p_5/\cos\theta)$.  Its Jacobian is
sometimes infinite, but not on the minimal solution for which 
minimization of  $H_1$ with respect to $p_5$ gives:
\begin{eqnarray}
\label{p5min}
p_5^{min} &=& \frac{-\cos\theta}
{\cos^2\theta/\chi_s + \sin^2\theta/\chi_{\pi}}
\int V_C(r-r^\prime) \rho(r^\prime) dr^\prime
\end{eqnarray}
After the variable change $H_1(\theta,q)$ can also be expanded in even
powers around the points $\theta = \pi/2, 3\pi/2$, so the minimal
solution does not cross these levels either. 
Altogether,  
$\theta_{min}(r)$ always stays in one of the four quadrants of the
circle.
For low dielectric constant, {\it i.e.} weak screening, we find that
phase separation is precluded altogether. The system exhibits a
homogeneous mixed phase in which the superspin points neither purely
in the AF or SC direction.  This state is reminiscent of the putative
supersolid phase in $^4$He as both order parameters (AF and SC) are
nonzero everywhere in the sample.

\begin{center}
\begin{figure}
\begin{center}
\resizebox{6.5cm}{4.8cm}{\includegraphics{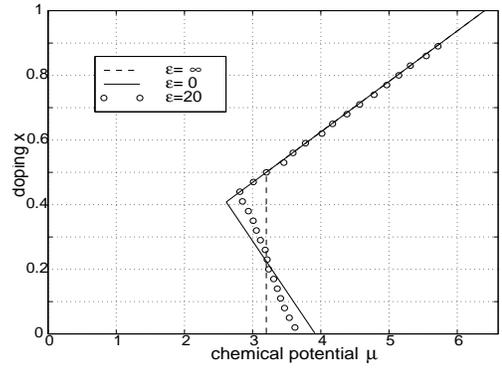}}
\end{center}
\vskip 0.1in
\caption{$x$ versus $\mu$ curve for three different values of $\epsilon$,
for paramteters $J=1$ and $K=0.4$ which yield $x_c=0.5$. 
The curves $\epsilon=0$ (homogeneous) and $\epsilon=\infty$  
(phase separation) are obtained from the analytic solution of Eq.~(\ref{xxz}). 
The circles are the data of the numerical solutions for $\epsilon=20$. }
\label{figure2}
\end{figure}
\end{center}

\begin{figure}[htbp]
\begin{center}
\resizebox{6.5cm}{4.0cm}{\includegraphics{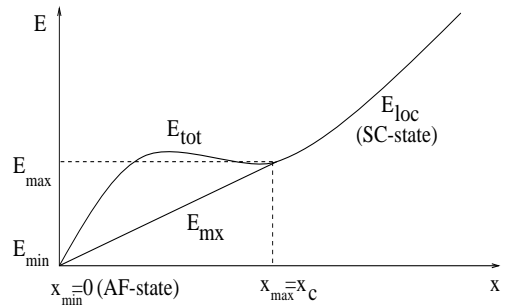}}
\end{center}
\caption{State energy as a function of $x$: 
In the phase separation regime, the energy becomes a convex function of $x$.}
\label{figuremu}
\end{figure}

After the minimum of $E$ is found, the chemical potential can be
numerically calculated as $\mu = \partial E/\partial x$.  As shown in
Fig.~\ref{figure2}, $\mu(x)$ becomes non-monotonic and a
region of $d\mu/dx < 0$ appears.  Such a region is prohibited in
thermodynamics, but in models with continuous charge density (as
opposed to point charges) it is generic. 
The origin of the region in which the chemical potential is double-valued
is illustrated by Fig.\ref{figuremu}.  In the absence of Coulomb
interactions and gradients the energy of a mixture with doping 
$x_{min}<x<x_{max}$ interpolates linearly between $E_{min}$ and
$E_{max}$.  The effect of Coulomb interactions and gradients is to
increase the total energy of these intermediate states to $E_{tot}(x)$
as shown in Fig.\ref{figuremu}.  The dependence of $\mu$ on $x$ is
then given simply by $\mu=\delta E_{tot}/\delta x$, which is consistent
with the numberical results of Fig.\ref{figure2}.

Because $E_{tot}(x_{min,max}) = E_{min,max}$ in models with continuous
charge one has
\beq \int_{x_{min}}^{x_{max}} (\mu(x)-\mu_c)\; dx =0.
\label{mu}
\eeq
which applies beyond the SO(5) theory. Our numerical results are
consistent with (\ref{mu}).

Experimentally, investigations of the chemical shifts in LSCO
\cite{ino} and BSCCO \cite{veeneedaal} have shown that while the 
shift is large in overdoped samples, it is strongly suppressed 
and pinned in underdoped samples, in agreement with the phase 
separation picture. However, due to poor experimental resolution, 
it is not possible to ascertain the non-monotonic behavior of $\mu(x)$. 
More experimental work is needed to test this prediction of our model.
While this work was motivated by experiment, it should be emphasized
that extensions of our model would be needed to make real contact with
experiments.  
The lattice anisotropy of the cuprates will lead to an anisotropic
AF/SC interface energy.  We expect this anisotropy to enlarge the
region of striped phase stability relative to that of the droplet
phases, as the former can take best advantage of that
anisotropy.
Also, disorder will make the
coefficients $J$ and $K$ (or $\chi_{ab}$ and $g$) and the charge of
the counterions position-dependent.  Although the resulting effects
are complex in character, we may speculate that for small disorder,
the defects act as pinning centers for the stripes and
lead to distortions of the domain structure as well as a loss of
long-range order. This may explain the failure to observe droplet
phases in the high-$T_c$ superconductors.  For strong randomness, the
size of the domains would be predominantly set by the disorder instead
of the long-range interactions \cite{tranquada1,chou}. However, we
expect that the linear relation between the superfluid density and the
doping should still hold in these glassy materials.

In summary, we have shown that the interplay between the long-range
Coulomb interaction and the local tendency to phase separation of the
SO(5) model leads to an interesting and remarkably rich phase diagram
for the clean system. We found that the frustrated phase separation
between hole-rich and hole-poor regions can provide an explanation for
the gross features of the cuprates near the AF/SC transition when
lattice anisotropy and impurity effects are taken into account. 
However, the SO(5) energy functional cannot have topological
solutions as its lowest energy state. 
Therefore we believe that the topological
nature of stripes that are observed in experiment must arise from
microscopic properties of the coexisting states.
Finally, we draw attention to
the behavior of the chemical potential in the phase separation regime.

We acknowledge many useful discussions with Eugene Demler and
Shoucheng Zhang.  This work was partially supported by NSERC, by the
Ontario Center for Materials Research and by Materials and
Manufacturing Ontario.


\begin{thebibliography}{10}
\bibitem{comment}
Modulated phases have also been obtained from mean-field treatments of the
  $t-J$ model \cite{shulz,zaanen,rice} as well as density matrix
  renormalization group calculations \cite{white}.
\bibitem{coul} Zhang's original theory\cite{zhang1} assumes that 
the short range part of the Coulomb interaction, the Hubbard $U$,
is the driving force for the strong correlations.  For homogeneous
systems the long range part is assumed to be unimportant.  The
situation is different for inhomogeneous systems for the reasons
discussed above.
\end{thebibliography}
\end{document}